\documentclass[a4paper,11pt]{article}
\pdfoutput=1 % if your are submitting a pdflatex (i.e. if you have
\usepackage{jcappub} % for details on the use of the package, please
\usepackage{graphicx}% Include figure files
\usepackage{dcolumn}% Align table columns on decimal point
\usepackage{bm}% bold math
\usepackage{physics}
\usepackage{xcolor}
\usepackage{comment}
\usepackage{amsmath}
\usepackage{mathrsfs}
\usepackage{tikz}

\usepackage{framed}
\definecolor{shadecolor}{gray}{0.85}

\newcommand{\bx}{\mathbf{x}}

\newcommand{\bk}{\mathbf{k}}

\newcommand{\md}{\mathrm{d}}

\title{\boldmath Imprints of flat space analyticity in de Sitter $\mathcal{S}$-matrix}

\author[1,2]{Jason Kristiano,}
\author[3,2]{Ryo Namba,}
\author[1,4,5,6]{Atsushi Naruko,}
\author[7,8]{Ryo Saito,}
\author[9]{Daisuke Yamauchi}

\affiliation[1]{Center for Gravitational Physics and Quantum Information, Yukawa Institute for Theoretical Physics, Kyoto University, Kyoto 606-8502, Japan}
\affiliation[2]{Research Center for the Early Universe (RESCEU), Graduate School of Science, The University of Tokyo, Tokyo 113-0033, Japan}
\affiliation[3]{RIKEN Center for Interdisciplinary Theoretical and Mathematical Sciences (iTHEMS), Wako, Saitama 351-0198, Japan}
\affiliation[4]{School of Natural Sciences, National Institute of Technology (KOSEN), Gunma College, 580 Toriba, Maebashi, Gunma 371-8530, Japan}
\affiliation[5]{Asia Pacific Center for Theoretical Physics, Pohang 37673, Korea}
\affiliation[6]{Department of Physics, Faculty of Engineering Science, Yokohama National University, Yokohama 240-8501, Japan}
\affiliation[7]{Graduate School of Science and Engineering, Yamaguchi University, Yamaguchi 753-8512, Japan}
\affiliation[8]{Kavli Institute for the Physics and Mathematics of the Universe, Todai Institute for Advanced Study,
The University of Tokyo, Chiba 277-8583, Japan (Kavli IPMU, WPI)}
\affiliation[9]{Department of Physics, Faculty of Science, Okayama University of Science, 1-1 Ridaicho, Okayama, 700-0005, Japan}

\emailAdd{jkristiano@yukawa.kyoto-u.ac.jp}
\emailAdd{ryo.namba@riken.jp}
\emailAdd{naruko@yukawa.kyoto-u.ac.jp}
\emailAdd{rsaito@yamaguchi-u.ac.jp}
\emailAdd{d-yamauchi@ous.ac.jp}

\abstract{
The analytic structure of the flat-space $\mathcal{S}$-matrix provides non-perturbative constraints on low-energy effective field theories based on the properties of high-energy theory. While the analytic structure of the flat-space $\mathcal{S}$-matrix is well understood, extending this framework to de Sitter space is challenging, as the expanding background complicates the definition of asymptotic states and breaks time-translation symmetry. This paper investigates how flat-space analyticity is imprinted on the de Sitter $\mathcal{S}$-matrix. We derive a relation between flat-space amplitude and de Sitter $\mathcal{S}$-matrix on a specific limit called the {\it Hubble flat-space} (HFS) limit. Specifically, we show that the relation holds for tree-level amplitude exchanging a massive scalar field with any local derivative interactions. Finally, we argue that the HFS limit is more compatible with the description of effective field theory, as the total energy dependence of de Sitter $\mathcal{S}$-matrix becomes trivial, allowing the Mandelstam variable to be identified as the unique energy scale, just as in flat space.
}

\begin{document}
\maketitle
\flushbottom

\section{Introduction}
The analytic structure of scattering amplitudes in flat spacetime has long played an important role in understanding quantum field theory (QFT) and perturbative quantum gravity \cite{Eden:1966dnq,Elvang:2015rqa}. With basic physical principles—unitarity, locality, Lorentz invariance, and causality—one finds that $\mathcal{S}$-matrix elements are not arbitrary functions of kinematic invariants but obey powerful constraints. Analyticity of amplitudes, in particular, encodes the causal propagation of intermediate states that connect high- and low-energy behavior. These features establish a link between infrared (IR) observables and ultraviolet (UV) theories \cite{Pham:1985cr, Adams:2006sv}.

In recent years, these ideas have found relevance in the context of the Swampland program, which seeks to characterize which low-energy effective field theories (EFTs) can arise from a consistent theory of quantum gravity. One of the key methods in this program is to identify universal constraints on EFTs that follow not from specific UV completions but from general properties of scattering amplitudes, such as analyticity and unitarity. These constraints, often formulated as positivity bounds or sum rules, link IR couplings to the behavior of amplitudes at high energies and have been used to exclude entire classes of EFTs from the ``landscape'' of consistent theories. In this way, the $\mathcal{S}$-matrix becomes a diagnostic of UV completeness --- even when specific details of the UV theory are unknown.

However, such a methodology relies on the existence of an $\mathcal{S}$-matrix, which assumes a globally defined notion of asymptotic states and energy eigenstates. This framework breaks down in cosmological spacetimes, such as de Sitter space, where the spacetime expansion complicates the definition of in/out states and the notion of particle propagation over long times \cite{Spradlin:2001nb, Bousso:2004tv, Marolf:2012kh, Dvali:2017eba, Albrychiewicz:2020ruh}. These challenges have led to a focus on equal-time correlation functions --- such as the cosmological wavefunction or late-time correlators --- as the primary observables in inflationary cosmology \cite{Maldacena:2002vr}. Recently, there has been rapid progress in bootstrapping cosmological correlators based on late-time de Sitter symmetry \cite{Mata:2012bx, Kundu:2014gxa, Kundu:2015xta, Ghosh:2014kba, Arkani-Hamed:2018kmz, Baumann:2019oyu, Baumann:2020dch}, which is essentially the conformal symmetry \cite{Bzowski:2013sza, Bzowski:2015pba, Bzowski:2018fql, Isono:2019ihz, Isono:2019wex}. Moreover, the bootstrap approach can be extended to capture a wider class of theories \cite{Chen:2006nt, Chen:2009we, Cheung:2007st, Cheung:2007sv, Creminelli:2006xe, Arkani-Hamed:2003juy} by relaxing the boost symmetry \cite{Pajer:2020wxk}. Also, the unitary and analytical properties of cosmological correlators have been studied in \cite{Cespedes:2020xqq, Jazayeri:2021fvk, DiPietro:2021sjt, Sleight:2021plv, Melville:2021lst, Goodhew:2021oqg, Albayrak:2023hie, AguiSalcedo:2023nds, Baumann:2021fxj, Ema:2024hkj, Salcedo:2022aal, Lee:2023kno, Liu:2024xyi}. While cosmological correlators carry important physical information, they do not naturally exhibit a similar analytic structure as the flat-space $\mathcal{S}$-matrix.

Recent works by Melville and Pimentel \cite{Melville:2023kgd, Melville:2024ove} offer a framework to bridge this gap. With the same spirit as the pioneering work by Marolf \textit{et al.} for global de Sitter \cite{Marolf:2012kh}, they construct a definition of a de Sitter $\mathcal{S}$-matrix for massive fields in the Poincar\'{e} patch, using coordinates $\md s^2 = (-H\tau)^{-2} (-\md\tau^2 + \md \bx^2)$, where the domain of the conformal time is $\tau \in (-\infty,0)$. Their approach defines transition amplitudes between asymptotic states --- either in the Bunch-Davies or Unruh-DeWitt vacuum by isolating the on-shell parts of time-ordered correlation functions and amputating external legs using appropriate mode functions. Importantly, this construction ensures that the resulting $\mathcal{S}$-matrix inherits invariance under field redefinitions and decoupling of total derivatives and remains insensitive to off-shell contact terms in the bulk action. 

Similarly, Donath and Pajer \cite{Donath:2024utn} define a de Sitter $\mathcal{S}$-matrix in the (double) Poincar\'{e} patch: the initial state of free particles at the past null horizon evolves in an interacting theory and projects the resulting state of free particles at the future null horizon. In this setup, the domain of the conformal time is extended to $\tau \in (-\infty, \infty)$. Both constructions of the de Sitter $\mathcal{S}$-matrix provide hope that some version of $\mathcal{S}$-matrix analyticity may be meaningful even in de Sitter space, and thus that one might extend the logic of flat-space UV/IR consistency into cosmology.

The goal of this paper is to explore precisely this possibility. The key question is: to what extent does the analytic structure of the flat-space $\mathcal{S}$-matrix leave imprints on the de Sitter $\mathcal{S}$-matrix? Specifically, we examine how known features --- such as the location and interpretation of singularities --- are deformed or preserved when scattering takes place in an expanding background with $\mathcal{S}$-matrix defined by Melville and Pimentel. This investigation is a first step toward a more ambitious question: can one formulate Swampland-type constraints on EFTs in de Sitter space using analytic properties of scattering amplitudes? In flat space, analyticity links low-energy data (e.g., the sign of a particular EFT operator coefficient) to assumptions about high-energy unitarity and causality. If an analogous formula can be developed in de Sitter, it may provide new tools to assess which inflationary models are compatible with a consistent UV embedding. In this way, the de Sitter $\mathcal{S}$-matrix could serve as a cosmological extension of the flat-space amplitude bootstrap, providing a method to assess properties of the UV theory from large-scale cosmological observables.

In this paper, we pursue this question by comparing explicit tree-level amplitudes in flat and de Sitter space. In section \ref{sec:summary}, we summarize the results of this paper. In section \ref{sec:smatrix}, we begin by reviewing the construction and properties of the de Sitter $\mathcal{S}$-matrix as defined in \cite{Melville:2023kgd, Melville:2024ove}, paying particular attention to its analytic structure. We then analyze specific examples: contact interactions and exchange diagrams. In section \ref{sec:conservation}, we first review the well-known energy conservation limit of the de Sitter $\mathcal{S}$-matrix. While the leading order contribution of this limit is well-studied, here we compute the next-to-leading order in which information from massive field exchange appears. In section \ref{sec:flatspace}, we introduce a new limit called the Hubble flat space (HFS) limit, where the information from massive field exchange appears at leading order. We also show a relationship between the flat-space and the de Sitter $\mathcal{S}$-matrix in this limit. In section \ref{sec:eft}, we show that the HFS limit provides a better description of the effective field theory compared to the energy conservation limit. We devote section \ref{sec:discussion} to discussion and future directions.

In this paper, we use the metric signature $(-,+,\dots,+)$, where the negative sign represents a time dimension and the positive sign represents $d$-spatial dimensions. We denote the $d$-dimensional spatial momentum vector and its magnitude of the $n$-th field as $\bk_n$ and $\abs{\bk_n} \equiv k_n$, respectively. For a massive scalar field in de Sitter, the mass is encoded in a dimensionless parameter $\mu^2 \equiv (m/H)^2 - (d/2)^2$. A special case called a conformally coupled scalar field has mass $i\mu = 1/2$. We are mainly interested in $2 \rightarrow 2$ scattering, so the index $n$ will be $n = 1,2,3,4$. Exchange momentum is denoted by $\bk_s = \bk_1 + \bk_2$ with a magnitude of $k_s \equiv \abs{\bk_s}$. The sum of the magnitudes of the $n$-th and $m$-th fields is denoted by $k_{nm} \equiv k_n + k_m$. A conformally coupled scalar field has an energy-momentum dispersion relation $\omega_n = \pm k_n$, where the $+$ and $-$ signs denote incoming and outgoing fields, respectively. For a $2 \rightarrow 2$ scattering of conformally coupled fields, $E \equiv -k_{12} + k_{34}$ is the total energy. In flat space QFT, given an $\mathcal{S}$-matrix, we define the scattering amplitude $\mathcal{M}$ by factoring out the conservation of energy and spatial momentum $\mathcal{S} \equiv i \mathcal{M} (2\pi)^{d+1} \delta(\Sigma_n \bk_n) \delta(\Sigma_n \omega_n)$. Similarly, in de Sitter, we can define the scattering amplitude $\mathcal{A}$ by factoring out the conservation of spatial momentum $\mathcal{S} \equiv \mathcal{A} (2\pi)^d \delta(\Sigma_n \bk_n)$. Later, we will define $\mathcal{A}'$ by factoring out the trivial normalization factor of the mode function from $\mathcal{A}$.

%%%%%%%%%%%%%%%%%%%%%%
%%
%%%%%%%%%%%%%%%%%%%%%%
\section{Summary}
\label{sec:summary}
It is known that the de Sitter $\mathcal{S}$-matrix or correlators do not satisfy energy conservation due to the background spacetime that breaks time-translation symmetry. 
This absence of symmetry introduces a new energy scale into the system: the total energy of the external legs of the diagram $\sum_a \omega_a$, which we denote by $E$. 
As a result, the de Sitter $\mathcal{S}$-matrix possesses a kinematic structure that is richer than its flat-space counterpart.

In flat space, the analytic structure of the $\mathcal{S}$-matrix is directly governed by the mass spectrum of a theory.
This naturally raises the question: how is the mass spectrum of a theory imprinted in the kinematic structure of the de Sitter $\mathcal{S}$-matrix?
As a partial answer, one may point to a characteristic feature of the de Sitter $\mathcal{S}$-matrix in {\it the energy conservation limit} \cite{Arkani-Hamed:2018kmz}:\footnote{This limit is referred to as the flat-space limit in Ref.~\cite{Arkani-Hamed:2018kmz}.}
\begin{shaded}
    \vspace*{0.1cm}
    \begin{equation}
        E \equiv \sum_a \omega_a \to 0 \,,
    \end{equation}
    \vspace*{-0.4cm}
\end{shaded}
\noindent
For example, consider a $\phi \phi \rightarrow \phi\phi$ scattering exchanging a massive field $\sigma$ with coupling $\phi^2\sigma$ in $d$ spatial dimensions. 
In the energy conservation limit, the de Sitter $\mathcal{S}$-matrix is proportional to
\begin{equation}
\lim_{E\rightarrow 0} \mathcal{A}_{2 \rightarrow 2} \propto \frac{1}{E^{d-4}} \mathcal{M}_{2 \rightarrow 2}^\mathrm{UV}. \label{econ}
\end{equation}
The residue of the total energy singularity, $\mathcal{M}_{2 \rightarrow 2}^\mathrm{UV}$, is the flat space scattering amplitude at high energy,  where the mass of the exchanged field is negligible compared to the external kinematics. 
Independently of the mass spectrum, it takes the universal form $\mathcal{M}_{2 \rightarrow 2}^\mathrm{UV} \propto 1/(s+ i \varepsilon)$ in the $s$-channel, where $s$ is the Mandelstam variable with an $i \varepsilon$-prescription. 
Thus, the energy conservation limit washes out information about the mass spectrum.

This motivates the following question: is there any limit in which the mass of the exchanged field appears at the leading order? 
To address this, we first examine the next-to-leading order correction in the energy conservation limit. We find that the $\mathcal{O}(E^2)$ correction is
\begin{align}
\lim_{E\rightarrow 0} \mathcal{A}_{2 \rightarrow 2} 
&\propto \frac{1}{E^{d-4}} \mathcal{M}_{2 \rightarrow 2}^\mathrm{UV} \left[ 1 + \mathcal{O}(1) \frac{m^2}{s} \left(\frac{E}{H}\right)^2 \right] ,
\end{align}
where $m$ is the mass of the exchanged field and $H$ is the Hubble parameter. 
Motivated by this structure, 
we introduce a new limit called the {\it Hubble flat-space} (HFS) limit defined as follows:
\begin{shaded}
    \vspace*{0.1cm}
    \begin{equation}
        E \to 0 \,,~ H \to 0 \,, \quad \text{while} \quad \alpha \equiv \frac{E}{H} = \text{finite}
    \end{equation}
    \vspace*{-0.4cm}
\end{shaded}
\noindent
Including all possible higher-derivative interactions with coefficients $c_n$, in this limit, we find the following relation between flat-space and de Sitter amplitude
\begin{align}
&\mathcal{A}_{2 \rightarrow 2}'(s;m,c_n) = (\sqrt{s})^{2-d} \frac{H}{2} \int_0^\infty \md\tilde{s} ~\tilde{s}^{\frac{d-4}{2}} e^{-i \alpha \sqrt{\tilde{s}/s}} \mathcal{M}_{2 \rightarrow 2}(\tilde{s};m ,c_n), \nonumber\\
&\mathcal{M}_{2 \rightarrow 2} (s; m, c_n) = \frac{-i}{s - m^2 + i\varepsilon} \sum_{n=0}^\infty c_n s^n.
\end{align}

A limit of similar spirit, called {\it the massive flat-space} (MFS) limit, has been introduced in  \cite{Cespedes:2025dnq}. 
In contrast to their construction, the HFS limit does not rely on the mass parameter $m$, and therefore applies equally well to theories without bare mass terms in the Lagrangian. 
Moreover, Ref.~\cite{Cespedes:2025dnq} focuses on equal time correlators, although some parts of the analytical derivation may look similar.

As it captures the mass spectrum of a theory, the HFS limit is more compatible for describing EFT than the energy conservation limit. In flat space QFT, we understand that a quartic contact amplitude is a low-energy limit of a cubic exchange amplitude, where low and high energy mean $s \ll m^2$ and $s \gg m^2$, respectively. We can identify $\sqrt{s}$ as the unique energy scale that is relevant for the EFT because the scattering amplitude has a nontrivial dependence on $s$, while the dependence on the total energy $E$ is simply $\delta(E)$. However, for the de Sitter amplitude in the energy conservation limit, a cubic exchange diagram scales as $\mathcal{O}(E^{4-d})$, while a quartic contact diagram scales as $\mathcal{O}(E^{2-d})$. In this limit, we cannot obtain an EFT description analogous to the flat space one. Moreover, it is not clear what constitutes the energy scale, since the amplitude has a nontrivial dependence on both $s$ and $E$. In contrast, in the HFS limit, we will show that the amplitude of the quartic contact and cubic exchange diagrams scales as $\mathcal{O}(E)$. Because the dependence on the total energy $E$ becomes trivial, we can identify $\sqrt{s}$ as the energy scale.

%%%%%%%%%%%%%%%%%%

\section{De Sitter $\mathcal{S}$-matrix}
\label{sec:smatrix}
In this section, we briefly review the de Sitter $\mathcal{S}$-matrix proposed by Melville and Pimentel \cite{Melville:2023kgd, Melville:2024ove}. The $\mathcal{S}$-matrix encodes how the ``in'' state of free particles at past infinity evolves into the ``out'' state of free particles at future infinity. To define an $\mathcal{S}$-matrix, one must specify a basis of the ``in'' and ``out'' states. While on Minkowski spacetime we have particle eigenstates $\ket{n}$ of the free theory as a natural choice of basis, on de Sitter space, we cannot do so because the number of particles is not conserved due to gravitational particle production. Because the Hamiltonian explicitly depends on time, the particle eigenstates have an explicit time dependence $\ket{n,\tau}$. Melville and Pimentel define a de Sitter $\mathcal{S}$-matrix by choosing the ``in'' and ``out'' states of the interacting theory to coincide with $\ket{n, -\infty}$ in the past and future infinity, respectively, that is given by
\begin{equation}
\mathcal{S}_{n \rightarrow n'} \equiv {}_\mathrm{out}\bra{n', -\infty}\ket{n, -\infty}_\mathrm{in}.
\end{equation}
They call this quantity the \textit{Bunch-Davies $\mathcal{S}$-matrix} because it describes the evolution of the Bunch-Davies vacuum state $\ket{0, -\infty}$ in the interacting theory.

As in Minkowski spacetime, the $\mathcal{S}$-matrix is computed within perturbation theory: quantize the free theory action $S_\mathrm{free}$, then include an interaction $S_\mathrm{int}$ as a small perturbation. For a real scalar field $\phi(\bx, \tau)$ with mass $m$ in $d$ spatial dimensions, the action reads
\begin{equation}
S_\mathrm{free} = \frac{1}{2} \int \md\tau ~\md^dx \sqrt{-g} \left[ - (\partial_\mu \phi)^2 - m^2 \phi^2 \right].
\end{equation}
As mentioned in the introduction, we consider the de Sitter metric in the Poincar\'{e} patch 
\begin{equation}
\md s^2 = \frac{1}{(-H\tau)^2} ( -\md \tau^2 + \md \bx^2 ),
\end{equation}
where $\tau$ is the conformal time with the domain $\tau \in (-\infty,0)$, and $\bx$ is the spatial coordinate. The action can be canonically normalized by introducing $\varphi \equiv (-H \tau)^{-d/2} \phi$. In Fourier space, the equation of motion for the free theory is
\begin{equation}
\mathcal{E}[k \tau] \varphi(\tau, \bk) \equiv \left[ (\tau \partial_\tau)^2 + (k \tau)^2 + \mu^2 \right] \varphi(\tau, \bk) = 0, \label{eom}
\end{equation}
where $\mu^2 \equiv (m/H)^2 - (d/2)^2$. The field $\varphi$ can be canonically quantized as
\begin{equation}
\hat{\varphi}(\tau, \bk) = f^-(k\tau) \, \hat{a}_{-\bk} + f^+(k\tau) \, \hat{a}_\bk^\dagger,
\end{equation}
where $f^\pm(k\tau)$ is the mode function satisfying $\mathcal{E}[k \tau] f^\pm(k\tau) = 0$. The positive frequency mode $f^+(k\tau)$ and the negative frequency mode $f^-(k\tau)$ are defined as solutions of the following equation
\begin{equation}
0 = \left. \left( \tau \partial_\tau \pm i \sqrt{k^2\tau^2 + \mu^2} \right) f^\pm(k\tau) \right|_{\tau = \tau_*},
\end{equation}
where $\tau_*$ is a reference time that depends on the choice of vacuum. For the Bunch-Davies $\mathcal{S}$-matrix, the vacuum condition is imposed at $\tau_* \rightarrow -\infty$, so the mode function is
\begin{equation}
f^+(k \tau) = \frac{\sqrt{\pi}}{2i \sqrt{H}} e^{\pi \mu/2} H_{i\mu}^{(2)} (-k\tau) = [f^-(k\tau)]^* ,
\end{equation}
where $H_{i\mu}^{(2)} (-k\tau)$ is the Hankel function of the second kind. The operators $\hat{a}_\bk^\dagger$ and $\hat{a}_\bk$ are the creation and annihilation operators, respectively, which satisfy the commutation relation $[\hat{a}_{\bk'}, \hat{a}_\bk^\dagger] = (2\pi)^d \delta(\bk' - \bk)$.

Analogous to QFT in Minkowski spacetime, the $\mathcal{S}$-matrix can be computed using the Lehmann-Symanzik-Zimmermann (LSZ) reduction formula \cite{Lehmann:1954rq}: starting from the time-ordered correlation function, amputate the external leg by applying the equation of motion, and then go on-shell by performing an integral transform using the mode function. These procedures read mathematically
\begin{align}
&\mathcal{S}_{n \rightarrow n'} = \prod_{b' = 1}^{n'} \int_{-\infty}^0 \frac{\md \tau_{b'}}{-\tau_{b'}} f^+(k_{b'} \tau_{b'}) \prod_{b = 1}^{n} \int_{-\infty}^0 \frac{\md \tau_{b}}{-\tau_{b}} f^-(k_{b} \tau_{b}) g_{n \rightarrow n'}, \nonumber\\
&g_{n \rightarrow n'} \equiv \prod_{b' = 1}^{n'} i \mathcal{E}[k_{b'} \tau_{b'}] \prod_{b = 1}^{n} i \mathcal{E}[k_{b} \tau_{b}] G_{n \rightarrow n'} ,\nonumber\\
&G_{n \rightarrow n'} \equiv {}_\mathrm{out}\bra{0} \mathrm{T}  \prod_{b' = 1}^{n'} \hat{\varphi}^\dagger(k_{b'} \tau_{b'}) \prod_{b = 1}^{n}  \hat{\varphi}(k_{b} \tau_{b}) \ket{0}_\mathrm{in}.
\end{align}
In this paper, the operator $\mathrm{T}$ denotes the time-ordered operator. With schematic notation, we introduce some terminology:
\begin{itemize}
    \item Correlator: ${}_\mathrm{out}\bra{0} \mathrm{T}  \hat{\varphi}_1  \hat{\varphi}_2 \dots \hat{\varphi}_n \ket{0}_\mathrm{in}$,
    \item Amputated correlator: $\left[ \prod_{j=1}^n i \mathcal{E}_j \right] {}_\mathrm{out}\bra{0} \mathrm{T}  \hat{\varphi}_1  \hat{\varphi}_2 \dots \hat{\varphi}_n \ket{0}_\mathrm{in}$,
    \item On-shell amputated correlator: $\left[ \prod_{j=1}^n \int_{\tau_j} f_j \right] \left[ \prod_{j=1}^n i \mathcal{E}_j \right] {}_\mathrm{out}\bra{0} \mathrm{T}  \hat{\varphi}_1  \hat{\varphi}_2 \dots \hat{\varphi}_n \ket{0}_\mathrm{in}$.
\end{itemize}
Based on the LSZ reduction formula, the $\mathcal{S}$-matrix can be computed from Feynman diagrams using the following rules:
\begin{itemize}
    \item Outgoing external lines: $f^+(k\tau)$,
    \item Ingoing external lines: $f^-(k\tau)$,
    \item Internal lines (propagator): $G(k \tau, k\tau')$,
    \item Vertex factor: $i \delta^n S_\mathrm{int}/\delta\varphi^n$,
    \item Integral over all internal times and momenta.
\end{itemize}
The propagator is defined in free theory two-point functions as \cite{Fukuma:2013mx}
\begin{equation}
\bra{0}\mathrm{T}\hat{\varphi}(\tau,\bk) \hat{\varphi}(\tau',\bk')\ket{0} \equiv G(k\tau, k\tau') (2\pi)^d \delta(\bk + \bk'),
\end{equation}
and in terms of the mode function, it is given by
\begin{equation}
G(k \tau, k\tau') = \theta(\tau - \tau' ) f^-(k\tau) f^+(k\tau') + \theta(\tau' - \tau) f^-(k\tau') f^+(k\tau).
\end{equation}
With these Feynman rules, we would like to compute the $\mathcal{S}$-matrix for simple examples of contact and exchange diagrams.

\subsection{Contact diagram}
In general, it is difficult to explicitly compute the de Sitter $\mathcal{S}$-matrix because of dealing with integrals of the Hankel function. However, technical computation can become simpler for some examples. For a specific case called conformally coupled scalar field, $i\mu = 1/2$, the mode function can be simplified to
\begin{equation}
f^\pm(k\tau) = \frac{1}{\sqrt{\mp 2i H k\tau}} e^{\pm i k\tau}.
\end{equation}
Consider a local interaction with Lagrangian
\begin{equation}
\label{eq:Lint_npoint}
\mathcal{L}_\mathrm{int} = \sqrt{-g} \frac{\lambda_n}{n!}  \phi^n.
\end{equation}
It generates an $n$-point contact contribution to the $\mathcal{S}$-matrix
\begin{align}
\mathcal{S}_{0 \rightarrow n} &= (2\pi)^d \delta\left( \sum_{a=1}^n \bk_a \right) i \lambda_n  \int_{-\infty}^0 \frac{\md \tau}{-\tau} (-H \tau)^{\frac{d}{2}(n-2)} \prod_{b = 1}^{n} f^+(k_{b} \tau)  \nonumber\\
&= (2\pi)^d \delta\left( \sum_{a=1}^n \bk_a \right) i \lambda_n  \int_{-\infty}^0 \frac{\md \tau}{-\tau} (-H \tau)^{\frac{d}{2}(n-2)- \frac{n}{2}} e^{i(k_1 + \dots + k_n) \tau} \prod_{b = 1}^{n} \frac{1}{\sqrt{2i k_b}} \nonumber\\
&= (2\pi)^d \delta\left( \sum_{a=1}^n \bk_a \right) i \lambda_n \Gamma(j_n) \left( \frac{H}{i E} \right)^{j_n} \prod_{b = 1}^{n} \frac{1}{\sqrt{2i k_b}},
\end{align}
where $E = k_1 + \dots + k_n$ is the total energy and 
\begin{equation}
j_n = \frac{n}{2}(d-1) -d.
\end{equation}
Hereafter, we will omit the momentum conservation factor $(2\pi)^d \delta\left( \sum_{a=1}^n \bk_a \right)$ and normalization $\prod_{b = 1}^{n} 1/\sqrt{2i k_b}$ and denote the resulting part of $\mathcal{S}$-matrix by $\mathcal{A}_{n \to n'}'$. Specifically, for $n=4$, it becomes
\begin{equation}
\mathcal{A}_{0 \rightarrow 4}' = i \lambda_4 \Gamma(d-2) \left( \frac{H}{i E} \right)^{d-2} .
\end{equation}
We implicitly implement the $i\varepsilon$-prescription
\begin{equation}
\int_{-\infty}^0 \md\tau ~e^{i(E - i\varepsilon)\tau} = \frac{1}{iE+\varepsilon}
\end{equation}
to ensure the convergence of the time integral. 

\subsection{Exchange diagram}
Another example, for $n=3$, the interaction \eqref{eq:Lint_npoint} also generates an $s$-channel four-point exchange contribution to the $\mathcal{S}$-matrix
\begin{align}
\mathcal{A}_{0 \rightarrow 4} = -\lambda_3^2 & \int_{-\infty}^0 \frac{\md \tau}{-\tau} (-H \tau)^{d/2} \int_{-\infty}^0 \frac{\md \tau'}{-\tau'} (-H \tau')^{d/2} \nonumber\\
&\times f^+(k_1 \tau) f^+(k_2 \tau) G(k_s \tau, k_s \tau') f^+(k_3 \tau') f^+(k_4 \tau'),
\end{align}
where $k_s \equiv \abs{\bk_1 + \bk_2}$ is the exchanged momentum and the momentum conservation factor is suppressed from displaying. With crossing symmetry, we can obtain $\phi(\bk_1) \phi(\bk_2) \rightarrow \phi(\bk_3) \phi(\bk_4)$ scattering amplitude
\begin{align}
\mathcal{A}_{2 \rightarrow 2} = -\lambda_3^2 &\int_{-\infty}^0 \frac{\md \tau}{-\tau} (-H \tau)^{d/2} \int_{-\infty}^0 \frac{\md \tau'}{-\tau'} (-H \tau')^{d/2} \nonumber\\
&\times f^-(k_1 \tau) f^-(k_2 \tau) G(k_s \tau, k_s \tau') f^+(k_3 \tau') f^+(k_4 \tau').
\end{align}
Substituting the mode function of the conformally coupled field, the $s$-channel becomes
\begin{equation}\label{eq:double-integral S}
\mathcal{A}_{2 \rightarrow 2}' = -\lambda_3^2 \int_{-\infty}^0 \frac{\md \tau}{-\tau} (-H \tau)^{\frac{d-2}{2}} \int_{-\infty}^0 \frac{\md \tau'}{-\tau'} (-H \tau')^{\frac{d-2}{2}} e^{-ik_{12}  \tau} G(k_s \tau, k_s \tau') e^{i k_{34} \tau'},
\end{equation}
where $k_{nm} \equiv k_n + k_m$. For $d=5$, the integral can be evaluated to
\begin{equation}
\label{eq:S_exchange}
\mathcal{A}_{2 \rightarrow 2}' = -\lambda_3^2 H^2 \frac{-k_{12} + k_{34} + 2k_s}{2k_s (-k_{12} + k_{34}) (-k_{12}  + k_s) ( k_{34} +k_s ) }.
\end{equation}
Total exchange contributions can be obtained by summing the $t$ and $u$-channel contributions.

\section{Energy conservation limit}
\label{sec:conservation}
In this paper, we are interested in the imprints of flat space amplitudes on the de Sitter $\mathcal{S}$-matrix. For QFT in Minkowski spacetime, the $\mathcal{S}$-matrix is proportional to $\delta(E)$, which means conservation of energy. However, in an expanding universe background, there is no time-translation symmetry; thus, energy during a scattering process does not generally conserve. Thus, it is reasonable to argue that the energy conservation limit of the de Sitter $\mathcal{S}$-matrix may contain information about flat space amplitudes. From its mode function, a conformally coupled field has a dispersion relation similar to that of a massless field on Minkowski spacetime that is $\omega = k$. For a scattering of conformally coupled fields $\phi(\bk_1) \phi(\bk_2) \rightarrow \phi(\bk_3) \phi(\bk_4)$, the total energy of the process is $E = -k_{12} + k_{34}$. 
In this section, we investigate the energy conservation limit of the de Sitter $\mathcal{S}$-matrix and demonstrate how the mass of the exchanged field emerges at the next-to-leading order.

\subsection{Conformally coupled field exchange}
In terms of total energy, the exchange contribution \eqref{eq:S_exchange} can be written as
\begin{equation}
\mathcal{A}_{2 \rightarrow 2}' = -\lambda_3^2 H^2 \frac{E + 2k_s}{2k_s E (E - k_{34}  + k_s) ( k_{34} +k_s ) } ,
\end{equation}
and taking energy conservation limit yields
\begin{equation}
\label{eq:S_exchange_ECL}
\lim_{E\rightarrow 0} \mathcal{A}_{2 \rightarrow 2}' = - \frac{1}{E} H^2 \frac{\lambda_3^2}{k_s^2 - k_{12}^2} = H^2 \frac{\lambda_3^2/s}{E} , 
\end{equation}
where $s \equiv k_{12}^2 - k_s^2$ is one of the Mandelstam variables. Checking this limit, we can deduce a simple relation: the residue of the total energy singularity in the de Sitter $\mathcal{S}$-matrix is the amplitude in Minkowski spacetime\footnote{For the de Sitter $\mathcal{S}$-matrix defined by Donath and Pajer \cite{Donath:2024utn}, instead of the total energy singularity, it possesses generalized energy conservation given by the derivative of the Dirac-delta function $\delta(E)$ \cite{Du:2025zpm, Du:2025glv}.}. This statement was first claimed in the context of AdS/CFT \cite{Raju:2012zs, Raju:2012zr} and then extended to de Sitter \cite{Arkani-Hamed:2017fdk, Baumann:2021fxj}.

While such a limit can be obtained from the final formula of $\mathcal{S}_{2 \rightarrow 2}$ after performing time integrals, we can show that conducting a series expansion in the time integral leads to the same limit. We redefine a time vertex $\tau' \rightarrow \tau + \Delta\tau$ so that the time integrals can be expressed as \cite{Cespedes:2025dnq}
\begin{align}
\mathcal{A}_{2 \rightarrow 2}' = -\lambda_3^2 H^{d-3} & \int_{-\infty}^\infty \md(\Delta\tau) \int_{-\infty}^0 \frac{\md \tau}{(-\tau)(-\tau - \Delta\tau)} (-\tau)^{\frac{d-2}{2}} (-\tau - \Delta\tau)^{\frac{d-2}{2}} \nonumber\\
& \times  e^{i E \tau} e^{ik_{34} \Delta\tau} G(k_s \tau, k_s (\tau+\Delta\tau)).
\end{align}
Performing the asymptotic expansion of the integrand as $\tau \rightarrow -\infty$ yields
\begin{equation}
\label{eq:S_exchange_taulimit}
\mathcal{A}_{2 \rightarrow 2}'  \simeq -\lambda_3^2 \int_{-\infty}^0 \frac{\md \tau}{(-\tau)^2} (-H\tau)^{d-2} e^{i E \tau} \int_{-\infty}^\infty \md(\Delta\tau)  e^{ik_{34} \Delta\tau} G(k_s \tau, k_s (\tau+\Delta\tau)).
\end{equation}
At an early time, the propagator has a simple formula
\begin{equation}
\lim_{\tau \rightarrow - \infty} G(k_s \tau, k_s (\tau+\Delta\tau)) = \frac{1}{2 k_s (-H \tau)} \left[ \theta(\Delta\tau)  e^{- i k_s \Delta\tau} + \theta(-\Delta\tau)  e^{i k_s \Delta\tau} \right].
\end{equation}
Integrating over $\Delta\tau$, the $\mathcal{S}$-matrix becomes
\begin{align}
\mathcal{A}_{2 \rightarrow 2}' & \simeq \lambda_3^2 H^{d-3} \int_{-\infty}^0 \md\tau (-\tau)^{d-5} e^{i E \tau} \frac{i}{2k_s} \left( \frac{1}{k_{34} + k_s - i\varepsilon} - \frac{1}{k_{34} - k_s + i \varepsilon} \right) \nonumber\\
&= \lambda_3^2 H^{d-3} \int_{-\infty}^0 \md\tau (-\tau)^{d-5} e^{i E \tau}  \frac{-i}{ k_{34}^2 - k_s^2 + i \varepsilon} \nonumber\\
&=  H^{d-3} \frac{\Gamma(d-4)}{(iE + \varepsilon)^{d-4}} \frac{i \lambda_3^2}{s + i \varepsilon} .
\end{align}
This gives the identical result with the energy conservation limit \eqref{eq:S_exchange_ECL} for $\varepsilon \to 0$ and $d=5$.
We see that the $\mathcal{S}$-matrix diverges when the spatial dimension $d$ is an integer less than or equal to four. To avoid dealing with this IR divergence and introducing a new scale associated with it, we simply focus on the case $d > 4$.

If we define the following quantity
\begin{equation}
\mathcal{M}_{2\rightarrow 2} (k_{34}, k_s; \tau) \equiv \lambda_3^2  \int_{-\infty}^\infty \md(\Delta\tau)  e^{ik_{34} \Delta\tau} (-\tau) G(k_s \tau, k_s (\tau+\Delta\tau)), \label{mdef}
\end{equation}
we can write the de Sitter $\mathcal{S}$-matrix \eqref{eq:S_exchange_taulimit} with $\tau \to -\infty$ in the integrand as
\begin{equation}
\lim_{E \rightarrow 0}\mathcal{A}_{2\rightarrow 2}' (E, k_{34}, k_s) = -H^{d-3} \int_{-\infty}^0 \md\tau (-\tau)^{d-5} e^{i E \tau} \mathcal{M}_{2\rightarrow 2} (k_{34}, k_s; \tau).
\end{equation}
For this simple example, that quantity is given by
\begin{equation}
\mathcal{M}_{2\rightarrow 2} (k_{34}, k_s; \tau) = \frac{i\lambda_3^2}{ k_{34}^2 - k_s^2 + i \varepsilon} = \frac{i\lambda_3^2}{s + i \varepsilon},
\end{equation}
which is the $\mathcal{S}$-matrix of a massless particle exchange in Minkowski spacetime. It provides a hint on the relation between the $\mathcal{S}$-matrix in Minkowski spacetime and de Sitter space. Then, a further question arises: how general is such a relation? If the scattering process exchanges a massive particle, is the relation still applicable?

\subsection{Massive particle exchange}
Consider a scattering of conformally coupled fields $\phi$ exchanging a scalar field $\sigma$ with mass $m_\sigma$. Their interaction is given by
\begin{equation}
\mathcal{L}_\mathrm{int} = \sqrt{-g} \frac{\lambda_3}{2}  \phi^2 \sigma.
\end{equation}
This interaction is intensively studied in the context of cosmological correlators, which is referred to as cosmological collider physics \cite{Arkani-Hamed:2015bza, Chen:2009we, Chen:2009zp, Noumi:2012vr}.

Back to scattering, such a cubic coupling generates an $s$-channel contribution to the de Sitter $\mathcal{S}$-matrix
\begin{equation}
\mathcal{A}_{2 \rightarrow 2}' = -\lambda_3^2 \int_{-\infty}^0 \frac{\md \tau}{-\tau} (-H\tau)^{\frac{d-2}{2}} \int_{-\infty}^0 \frac{\md \tau'}{-\tau'} (-H\tau')^{\frac{d-2}{2}} e^{-ik_{12}  \tau} G_\sigma(k_s \tau, k_s \tau') e^{i k_{34} \tau'}, \label{start}
\end{equation}
where $G_\sigma$ is the propagator of the $\sigma$ field. To obtain the energy conservation limit, we follow the same strategy as in the previous subsection: redefine the integral variable and expand the integrand at $\tau \rightarrow -\infty$.

At an early time, the leading order expansion of the propagator is
\begin{equation}
\lim_{\tau \rightarrow - \infty} G_\sigma(k_s \tau, k_s (\tau+\Delta\tau)) = \frac{1}{2 k_s (-H \tau)} \left[ \theta(\Delta\tau)  e^{- i k_s \Delta\tau} + \theta(-\Delta\tau)  e^{i k_s \Delta\tau} \right],
\end{equation}
which is equivalent to the conformally coupled case. To obtain the dependence on mass, we have to expand it to the next leading order
\begin{align}
&\lim_{\tau \rightarrow - \infty} G_\sigma(k_s \tau, k_s (\tau+\Delta\tau))  \nonumber\\
&= \theta(\Delta\tau) \frac{e^{-i k_s \Delta\tau}}{16H (-k_s\tau)^3} \left[ -1 + 3 k_s^2 (\Delta\tau)^2 + 8 k_s^2 \tau^2 - 4 \mu_\sigma^2 - k_s \Delta\tau (i + 4 k_s \tau+ 4i \mu_\sigma^2) \right] \nonumber\\
&+ \theta(-\Delta\tau) \frac{e^{i k_s \Delta\tau}}{16H(-k_s\tau)^3} \left[ -1 + 3 k_s^2 (\Delta\tau)^2 + 8 k_s^2 \tau^2 - 4 \mu_\sigma^2 - k_s \Delta\tau (-i + 4 k_s \tau - 4i \mu_\sigma^2) \right], \label{asym}
\end{align}
where $\mu_\sigma^2 \equiv (m_\sigma/H)^2 - (d/2)^2$ is defined. Isolating dependence on $\mu_\sigma$ or considering the heavy mass limit $\mu_\sigma \gg 1$, the propagator becomes
\begin{align}
\lim_{\substack{\tau \rightarrow - \infty \\ \mu_\sigma \gg 1}} G_\sigma(k_s \tau, k_s (\tau+\Delta\tau)) = & ~\theta(\Delta\tau)\frac{e^{-i k_s \Delta\tau}}{16H (-k_s\tau)^3} \left[ 8 k_s^2 \tau^2 - 4 \mu_\sigma^2 (1 + i k_s \Delta\tau) \right] \nonumber\\
& + \theta(-\Delta\tau) \frac{e^{i k_s \Delta\tau}}{16H (-k_s\tau)^3} \left[ 8 k_s^2 \tau^2 - 4 \mu_\sigma^2 (1 - i k_s \Delta\tau) \right].
\end{align}
Substituting it into the de Sitter $\mathcal{S}$-matrix and identifying the limit with the energy conservation one yields
\begin{align}
\lim_{E \rightarrow 0} \mathcal{A}_{2 \rightarrow 2}' &= \lambda_3^2 H^{d-3} \int_{-\infty}^0 \md\tau (-\tau)^{d-5} e^{i E \tau} (-i) \left[ \frac{1}{ k_{34}^2 - k_s^2 } + \frac{\mu_\sigma^2}{(-\tau)^2 (k_{34}^2 - k_s^2)^2} \right] \nonumber\\
&= H^{d-3} \frac{\Gamma(d-4)}{(iE + \varepsilon)^{d-4}} \frac{i \lambda_3^2}{s + i \varepsilon} \left[ 1 + \frac{\Gamma(d-6)}{\Gamma(d-4)} (i E)^2 \frac{\mu_\sigma^2}{s} \right]. \label{end}
\end{align}

From this next leading order expansion, we can improve the previous statement on the residue of total energy singularity in de Sitter $\mathcal{S}$-matrix. It turns out that the residue of the leading singularity is the high-energy amplitude of scattering in Minkowski spacetime, where the exchanged field appears massless. Moreover, the mass of the exchanged particle appears as the residue of the next leading order singularity.

\section{Hubble flat-space analyticity}
\label{sec:flatspace}
In flat space QFT, we understand that a scattering process exchanging a massive scalar field yields an amplitude
\begin{equation}
\mathcal{M}_{2\rightarrow 2} (s;m) \propto \frac{i}{s - m^2 + i \varepsilon},
\end{equation}
which shows a singular behavior as $s$ approaches $m^2$. This is one of the analytical properties of the flat-space $\mathcal{S}$-matrix. However, as discussed in the final part of the previous section, it is difficult to see how this analytical property is manifested in de Sitter $\mathcal{S}$-matrix. The mass of the exchanged field appears as the residue of a higher-order singularity. An important question is, can we have a limit where $s$ and $m^2$ appear on the same footing? In this section, in contrast to the energy conservation limit $E \rightarrow 0$, we would like to discuss a limit called the {\it Hubble flat-space (HFS)} limit: $E \rightarrow 0$ and $H \rightarrow 0$, while keeping the ratio $\alpha \equiv E/H$ finite. We will compute de Sitter $\mathcal{S}$-matrix of conformally coupled fields exchanging a massive field in the HFS limit using integral and differential equation approaches. In the integral approach, we directly compute the time integral by performing an approximation to the integrand. In the differential equation approach, we find an exact differential equation of the $\mathcal{S}$-matrix as function of kinematical variables, and then approximate the differential equation to the HFS limit. 

\subsection{Integral approach}
In this subsection, we will re-do the computation from \eqref{start} to \eqref{end} in the HFS limit. This time, we have to carefully check whether the asymptotic expansion \eqref{asym} is valid in such a limit. We start from the integral representation of the Hankel function
\begin{equation}
f^\pm (k\tau) = - \frac{\sqrt{\pi}}{\sqrt{H}} \int_{-\infty}^\infty \frac{\md\rho}{2\pi} e^{i\mu\rho} e^{\pm i k\tau \cosh\rho}.
\end{equation}
As a simple check, in the limit $k\tau \rightarrow - \infty$, we can use the saddle-point approximation
\begin{equation}
\int_{-\infty}^\infty \md \rho ~F(\rho) e^{\pm i k\tau \cosh\rho} = F(0) e^{i\pi/4} \sqrt{\frac{2\pi}{\pm k\tau}} \, e^{\pm i k \tau}
\end{equation}
to obtain the early time limit of the mode function
\begin{equation}
f^\pm (k\tau \rightarrow - \infty) = \frac{1}{\sqrt{\mp 2i H k\tau}} e^{\pm i k\tau}.
\end{equation}
Pushing the time argument in the propagator to the early time yields
\begin{align}
&\lim_{\tau \rightarrow - \infty} G_\sigma(k \tau, k (\tau+\Delta\tau))= - \frac{\sqrt{\pi}}{H} \int_{-\infty}^\infty \frac{\md\rho}{2\pi} e^{i \mu_\sigma \rho} \frac{1}{\sqrt{-2k\tau}} \nonumber\\
&\times \left[ \theta(\Delta\tau) \frac{1}{\sqrt{-i}} e^{i k\tau} e^{-i k(\tau + \Delta\tau)\cosh\rho } + \theta(-\Delta\tau) \frac{1}{\sqrt{i}} e^{-i k\tau} e^{i k (\tau + \Delta\tau) \cosh\rho} \right]. \label{gsigma}
\end{align}
At this point, we have not performed any asymptotic expansion on the propagator. We only evaluate the time argument at $\tau \rightarrow - \infty$.

After performing the technical computation shown in Appendix \ref{saddle}, we obtain
\begin{align}
\mathcal{M}_{2\rightarrow 2} &= \frac{i}{k_{34}^2 - k_s^2 - (\mu_\sigma/\tau)^2 + i \varepsilon} = \frac{i}{s - (\mu_\sigma/\tau)^2 + i \varepsilon} \ , \label{mres} \\
\mathcal{A}_{2 \rightarrow 2}' &= H^{d-3} \int_{-\infty}^0 \md\tau (-\tau)^{d-5} e^{i E \tau} \frac{-i}{s - (\mu_\sigma/\tau)^2 + i \varepsilon} \nonumber\\
& = \alpha^{3-d} E \int_0^\infty \md\xi ~\xi^{d-3} e^{-i\xi} \frac{-i}{s \xi^2 - \mu_\sigma^2 E^2 + i\varepsilon} \ . \label{int_flat}
\end{align}
With the Sokhotski-Plemelj theorem, the integrand can be expressed as
\begin{equation}
\frac{1}{s \xi^2 - \mu_\sigma^2 E^2 + i\varepsilon} = -i \pi \delta\left( s \xi^2 - \mu_\sigma^2 E^2 \right) + \mathcal{P} \left( \frac{1}{s \xi^2 - \mu_\sigma^2 E^2 }  \right).
\end{equation}
As an example, for $d=5$, we can exactly evaluate the integral
\begin{equation}
\mathcal{A}_{2 \rightarrow 2}' = \frac{E}{\alpha^2 s} \lambda_3^2 \left\lbrace -\frac{i}{2} \xi_c \left[ e^{i \xi_c} \mathrm{Ei}(-i \xi_c) - e^{-i \xi_c} \mathrm{Ei}(i \xi_c)  \right] - 1 + i \pi \xi_c \sin \xi_c - \frac{\pi}{2} \xi_c e^{-i \xi_c} \right\rbrace, \label{sold5}
\end{equation}
where $\xi_c$ is the pole given by
\begin{equation}
\xi_c \equiv \frac{\mu_\sigma E}{\sqrt{s}} = \frac{m_\sigma}{\sqrt{s}} \alpha.
\end{equation}

\subsection{Differential equation approach}
In the previous subsection, we computed the integral using the saddle-point approximation. In this subsection, as a cross-check, we want to compute the same integral indirectly by obtaining a differential equation in general, then performing the limit. The de Sitter $\mathcal{S}$-matrix \eqref{start} can be written as
\begin{equation}
\mathcal{A}_{2 \rightarrow 2}' = -\lambda_3^2 \left( \frac{H}{k_s} \right)^{d-2} \int_{-\infty}^0 \frac{\md z}{z} (-z)^{\frac{d-2}{2}} \int_{-\infty}^0 \frac{\md z'}{z'} (-z')^{\frac{d-2}{2}} e^{-i z k_{12}/k_s} G_\sigma(z,z') e^{i z' k_{34}/k_s}.
\end{equation}
In Appendix \ref{diffeq}, we show that $\mathcal{A}_{2 \rightarrow 2}'$ satisfies the following differential equation
\begin{equation}
\left[ (k_{12}^2 - k_s^2) \partial_{k_{12}}^2 + (d-1) k_{12} \partial_{k_{12}} + \mu_\sigma^2 + \frac{(d-2)^2}{4} \right] \mathcal{A}_{2 \rightarrow 2}'  = i \Gamma(d-2) \frac{\lambda_3^2}{H} \left( \frac{H}{i E} \right)^{d-2}.
\end{equation} 
In the HFS limit, it reduces to
\begin{equation}
\left( s \partial_E^2 + \mu_\sigma^2 \right) \mathcal{A}_{2\rightarrow 2}' = i \Gamma(d-2) \frac{\lambda_3^2}{H} \left( \frac{H}{iE} \right)^{d-2}.
\end{equation}
We can solve the differential equation by substituting the ansatz
\begin{equation}
\mathcal{A}_{2\rightarrow 2}' \equiv \int_{-\infty}^0 \md \tau \mathscr{S}(\tau) e^{iE\tau},
\end{equation}
so we obtain the function $\mathscr{S}(\tau)$ as
\begin{align}
(-s\tau^2 + \mu_\sigma^2) \mathscr{S}(\tau) &= i \lambda_3^2 (-H\tau)^{d-3} \nonumber\\
\mathscr{S}(\tau) &= (-H\tau)^{d-3}  \frac{ - i \lambda_3^2 }{s\tau^2 - \mu_\sigma^2}.
\end{align}
Thus, we obtain the same solution as the previous approach \eqref{int_flat}.

Such a differential equation is studied in the context of the cosmological collider \cite{Arkani-Hamed:2015bza} and the cosmological bootstrap \cite{Arkani-Hamed:2018kmz, Baumann:2019oyu, Baumann:2020dch}. In this approach, it is straightforward to include all possible higher-derivative contact interactions
\begin{equation}
\left( s \partial_E^2 + \mu_\sigma^2 \right) \mathcal{A}_{2\rightarrow 2}' = \frac{i}{H} \left( \frac{H}{iE} \right)^{d-2} \sum_{n=0}^\infty c_n \Gamma(d-2+ 2n)  \left( \frac{H}{E} \right)^{2n}s^n.
\end{equation}
The solution is given by
\begin{align}
\mathcal{A}_{2 \rightarrow 2}' & = H^{d-3} \int_{-\infty}^0 \md\tau (-\tau)^{d-5} e^{i E \tau} \frac{-i}{s - (\mu_\sigma/\tau)^2 + i \varepsilon} \sum_{n=0}^\infty c_n \left( s (-H\tau)^2 \right)^n \nonumber\\
& = \alpha^{3-d} E \int_0^\infty \md\xi ~\xi^{d-3} e^{-i\xi} \frac{-i}{s \xi^2 - \mu_\sigma^2 E^2 + i\varepsilon} \sum_{n=0}^\infty \frac{c_n}{\alpha^{2n}} \left( s \xi^2 \right)^n.
\end{align}
If we define $\tilde{s} \equiv s (\xi/\alpha)^2$, we can write the solution as
\begin{equation}
\mathcal{A}_{2 \rightarrow 2}' = (\sqrt{s})^{2-d} \frac{H}{2} \int_0^\infty \md\tilde{s} ~\tilde{s}^{\frac{d-4}{2}} e^{-i \alpha \sqrt{\tilde{s}/s}} \frac{-i}{\tilde{s} - m_\sigma^2  + i\varepsilon} \sum_{n=0}^\infty c_n  \tilde{s}^{n}. 
\end{equation}
We can read that the integrand is proportional to the most general exchange amplitude in flat space, which includes all possible higher-derivative couplings. Thus, we can deduce a relation between flat space and de Sitter $\mathcal{S}$-matrix
\begin{equation}
\mathcal{A}_{2 \rightarrow 2}'(s;m,c_n) = (\sqrt{s})^{2-d} \frac{H}{2} \int_0^\infty \md\tilde{s} ~\tilde{s}^{\frac{d-4}{2}} e^{-i \alpha \sqrt{\tilde{s}/s}} \mathcal{M}_{2 \rightarrow 2}(\tilde{s};m ,c_n). \label{dsflat}
\end{equation}

\section{Effective field theory}
\label{sec:eft}
In flat space QFT, we understand that the scattering amplitude with a massive particle exchange reduces to a contact amplitude at low energy. This is an example of effective field theory. Consider a theory of a massless field $\phi$ and a massive field $\sigma$ with a Lagrangian 
\begin{equation}
\mathcal{L} = \sqrt{-g} \left[ - \frac{1}{2} (\partial_\mu \phi)^2 - \frac{1}{2} (\partial_\mu \sigma)^2 - \frac{1}{2} m^2 \sigma^2 + \frac{\lambda_3}{2}  \phi^2 \sigma \right].
\end{equation}
The cubic coupling induces an exchange amplitude
\begin{equation}
\mathcal{M}_{2 \rightarrow 2} = -i \lambda_3^2 \left( \frac{1}{s - m^2} + \frac{1}{t - m^2} + \frac{1}{u - m^2} \right).
\end{equation}
At low energy, where $s,t,u \ll m^2$, the exchange amplitude reduces to
\begin{equation}
\mathcal{M}_{2 \rightarrow 2} \simeq 3i \lambda_3^2 \frac{1}{m^2} ,
\end{equation}
which can be regarded as a contact amplitude induced by a quartic coupling of $\phi$ without any derivatives.

We can show that the cubic coupling becomes a quartic coupling in the Lagrangian by substituting the equation of motion for the massive field. This procedure can be considered as integrating out the heavy field. The equation of motion for the massive field is given by
\begin{align}
(\Box - m^2) \sigma &= - \frac{\lambda_3}{2} \phi^2 \nonumber\\
\sigma &= - \frac{\lambda_3}{2} \frac{1}{\Box - m^2} \phi^2,
\end{align}
where the inverse of an operator is a schematic notation. Performing a low-energy expansion
\begin{equation}
\frac{1}{\Box - m^2} \simeq -\frac{1}{m^2} - \frac{\Box}{m^4} + \dots,
\end{equation}
and substituting it into the Lagrangian yields
\begin{equation}
\mathcal{L} = - \frac{1}{2} (\partial_\mu \phi)^2 + \frac{\lambda_3^2}{8}\phi^2 \frac{1}{\Box - m^2} \phi^2 \simeq - \frac{1}{2} (\partial_\mu \phi)^2 - \frac{\lambda_3^2}{8m^2} \phi^4 + \frac{\lambda_3^2}{8m^4} \phi^2 \Box \phi^2 + \dots
\end{equation}
We can see that the coefficient of the lowest order quartic coupling is $\lambda_4 \equiv - \lambda_3^2/(8m^2)$. The schematic representation of the effective field theory can be summarized as
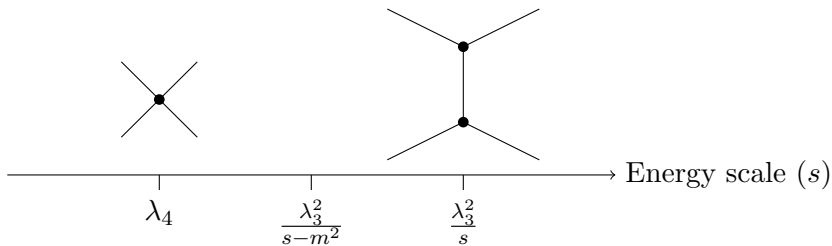
\newline
\newline
\begin{tikzpicture}
  % Draw axis
  \draw[->] (0,0) -- (8,0) node[anchor=west] {Energy scale $(s)$};
  % Points on the axis
  \draw (2,0) -- (2,-0.2) node[below] {$\lambda_4$};
  \draw (4,0) -- (4,-0.2) node[below] {$\frac{\lambda_3^2}{s-m^2}$};
  \draw (6,0) -- (6,-0.2) node[below] {$\frac{\lambda_3^2}{s}$};

\begin{scope}[shift={(2,1)}]
    % Incoming lines
    \draw[-] (-0.5,0.5) -- (0,0);
    \draw[-] (-0.5,-0.5) -- (0,0);
    % Outgoing lines
    \draw[-] (0,0) -- (0.5,0.5);
    \draw[-] (0,0) -- (0.5,-0.5);
    % Contact vertex
    \fill (0,0) circle (2pt);
  \end{scope}

  \begin{scope}[shift={(6,1.2)}]
    % Incoming lines
    \draw[-] (-1,1) -- (0,0.5);
    \draw[-] (-1,-1) -- (0,-0.5);
    % Propagator (vertical line)
    \draw (0,0.5) -- (0,-0.5);
    % Outgoing lines
    \draw[-] (0,0.5) -- (1,1);
    \draw[-] (0,-0.5) -- (1,-1);
    % Vertices
    \fill (0,0.5) circle (2pt);
    \fill (0,-0.5) circle (2pt);
  \end{scope}
\end{tikzpicture}

In flat space QFT, when we consider a specific channel, for example the $s$-channel, Lorentz invariance requires the kinematics of the scattering amplitude to be a function of $s$, the Mandelstam variable. For the scattering considered above, high and low energy are clear: $s \gg m^2$ and $s \ll m^2$, respectively. The limits are obvious because the Mandelstam variable $s$ is the unique kinematics in the scattering amplitude.

However, the situation is less clear in the de Sitter $\mathcal{S}$-matrix. Even in the energy conservation limit, for a specific channel, it cannot be reduced to a single kinematical dependence. It was shown in \eqref{end} that the $s$-channel contribution is a function of the Mandelstam variable $s$ and the total energy $E$. While the dependence on the total energy is trivial for scattering amplitude on flat space that is $\delta(E)$, the de Sitter $\mathcal{S}$-matrix has a non-trivial dependence on the total energy even at $E \rightarrow 0$ limit: from $\mathcal{O}(E^{4-d})$ to $\mathcal{O}(E^{2-d})$. In the energy conservation limit, a naive schematic picture of the effective field theory is
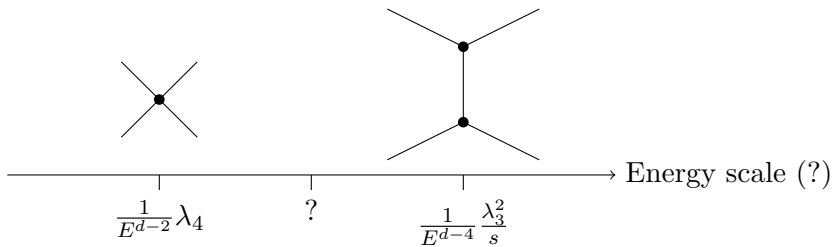
\newline
\newline
\begin{tikzpicture}
  % Draw axis
  \draw[->] (0,0) -- (8,0) node[anchor=west] {Energy scale $(?)$};

  % Points on the axis
  \draw (2,0) -- (2,-0.2) node[below] {$\frac{1}{E^{d-2}} \lambda_4$};
  \draw (4,0) -- (4,-0.2) node[below] {$?$};
  \draw (6,0) -- (6,-0.2) node[below] {$\frac{1}{E^{d-4}} \frac{\lambda_3^2}{s}$};

\begin{scope}[shift={(2,1)}]
    % Incoming lines
    \draw[-] (-0.5,0.5) -- (0,0);
    \draw[-] (-0.5,-0.5) -- (0,0);
    % Outgoing lines
    \draw[-] (0,0) -- (0.5,0.5);
    \draw[-] (0,0) -- (0.5,-0.5);
    % Contact vertex
    \fill (0,0) circle (2pt);
  \end{scope}

  \begin{scope}[shift={(6,1.2)}]
    % Incoming lines
    \draw[-] (-1,1) -- (0,0.5);
    \draw[-] (-1,-1) -- (0,-0.5);
    % Propagator (vertical line)
    \draw (0,0.5) -- (0,-0.5);
    % Outgoing lines
    \draw[-] (0,0.5) -- (1,1);
    \draw[-] (0,-0.5) -- (1,-1);
    % Vertices
    \fill (0,0.5) circle (2pt);
    \fill (0,-0.5) circle (2pt);
  \end{scope}
\end{tikzpicture}
\newline
In this picture, it is unclear what should be considered the energy scale to be compared with the mass of the heavy field $m$ because the kinematical dependence does not reduce to a unique variable. One might argue that it is sufficient to consider $s$ as the energy scale; however, the exchange and contact diagrams have a different dependence on the total energy $E$.

Then, how about the HFS limit? In this limit, the schematic picture becomes
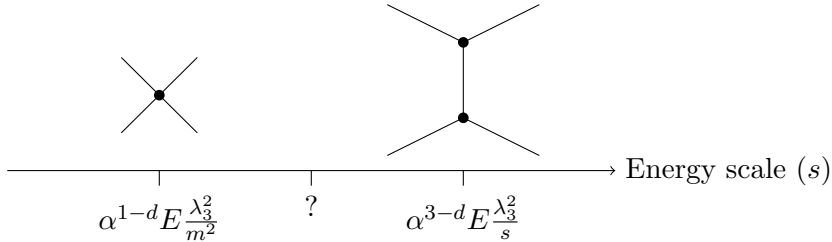
\newline
\newline
\begin{tikzpicture}
  % Draw axis
  \draw[->] (0,0) -- (8,0) node[anchor=west] {Energy scale $(s)$};

  % Points on the axis
  \draw (2,0) -- (2,-0.2) node[below] {$\alpha^{1-d} E \frac{\lambda_3^2}{m^2}$};
  \draw (4,0) -- (4,-0.2) node[below] {$?$};
  \draw (6,0) -- (6,-0.2) node[below] {$\alpha^{3-d} E \frac{\lambda_3^2}{s}$};

\begin{scope}[shift={(2,1)}]
    % Incoming lines
    \draw[-] (-0.5,0.5) -- (0,0);
    \draw[-] (-0.5,-0.5) -- (0,0);
    % Outgoing lines
    \draw[-] (0,0) -- (0.5,0.5);
    \draw[-] (0,0) -- (0.5,-0.5);
    % Contact vertex
    \fill (0,0) circle (2pt);
  \end{scope}

  \begin{scope}[shift={(6,1.2)}]
    % Incoming lines
    \draw[-] (-1,1) -- (0,0.5);
    \draw[-] (-1,-1) -- (0,-0.5);
    % Propagator (vertical line)
    \draw (0,0.5) -- (0,-0.5);
    % Outgoing lines
    \draw[-] (0,0.5) -- (1,1);
    \draw[-] (0,-0.5) -- (1,-1);
    % Vertices
    \fill (0,0.5) circle (2pt);
    \fill (0,-0.5) circle (2pt);
  \end{scope}
\end{tikzpicture}
\newline
In the intermediate scale, referring to \eqref{int_flat}, the question mark is given by
\begin{equation}
\alpha^{3-d} E \int_0^\infty \md\xi ~\xi^{d-3} e^{-i\xi} \frac{-i}{s \xi^2 - \alpha^2 m^2 + i\varepsilon}.
\end{equation}
In the HFS limit, we can attribute the energy scale to the Mandelstam variable $s$ because the dependence on the total energy trivializes to $\mathcal{O}(E)$ from high to low energy. For $d=5$, the solution is given by \eqref{sold5}, where the high and low energy limits are $\xi_c \rightarrow 0$ and $\xi_c \rightarrow \infty$, respectively \cite{DuasoPueyo:2025lmq}.

\section{Discussion and future directions}
\label{sec:discussion}
In this paper, we have established a new limit, called the {\it Hubble flat-space} (HFS) limit, for the de Sitter $\mathcal{S}$-matrix by taking $E \rightarrow 0$ and $H \rightarrow 0$ while keeping the ratio $\alpha \equiv E/H$ finite. In this limit, the de Sitter $\mathcal{S}$-matrix can be expressed as a specific integral transform of its flat space counterpart. The previously studied energy conservation limit ($E \rightarrow 0$ with $H$ fixed) only recovers the massless, high-energy limit of the flat-space amplitude. The HFS limit has the advantage that it can recover the full tree-level flat space amplitude, including its dependence on the mass $m$ of the exchange particle. Moreover, this limit is more compatible with the notion of EFT. By trivializing the total energy dependence to $\mathcal{O}(E)$ for both contact and exchange diagrams, the Mandelstam variable $s$ becomes the unique and relevant energy scale for scattering.

Some open questions for future research:
\begin{itemize}
    \item Deriving de Sitter constraints from (Hubble) flat-space analyticity: An important future task is to use the integral relation \eqref{dsflat}. Our central motivation is to see if flat-space analyticity and positivity constraints, which are foundational to the Swampland program, could be imported into de Sitter.
    \item Generalization to spinning particles: Our analysis was restricted to massive scalar fields. A crucial generalization will be to include spinning particles, most importantly, the graviton. Extending our integral relation to include graviton exchange would be an important step toward a non-perturbative understanding of quantum gravity in de Sitter. This is technically challenging but essential for making contact with Swampland conjectures that are fundamentally about gravity.
    \item Beyond tree-level: Our results are strictly tree-level. The flat-space constraints we hope to import (particularly those from unitarity) are non-perturbative. The next question is how our integral relation is modified by loop corrections. Understanding the (Hubble) flat-space limit of the de Sitter $\mathcal{S}$-matrix at loop level, and whether it can be related to its flat space amplitude counterpart in a similar way, is important for future work.
    \item Application to inflationary cosmology: In the EFT of inflation, the leading cubic couplings are given by de Sitter breaking interactions \cite{Chen:2006nt, Chen:2009we, Cheung:2007st, Cheung:2007sv, Creminelli:2006xe, Arkani-Hamed:2003juy}, which are $\dot{\phi}^3$ and $\dot{\phi}(\partial_i \phi)^2$. It is interesting to analyze the analytical properties of the $\mathcal{S}$-matrix involving those interactions in a de Sitter background. The goal is to obtain a positivity constraint on the coefficients of such de Sitter breaking interactions, which has been studied in \cite{Baumann:2015nta, Kim:2019wjo, Ye:2019oxx, Grall:2021xxm} to some extent.
\end{itemize}

\acknowledgments
J.K. thanks Katsuki Aoki and Andrea Cristofoli for the discussion. J.K. is supported by JSPS KAKENHI Grants Number 
JP25K23380. This work 
was partly supported by JSPS KAKENHI Grant Numbers JP23H01171 (J.K., R.N., A.N., R.S., and D.Y.), JP23K25868 (J.K., R.N., A.N., R.S., and D.Y.), JP20H05852(A.N.) and JP22K03627 (D.Y.). 

\appendix

\section{Saddle-point approximation \label{saddle}}

In this appendix, we show the technical derivation of the massive field exchange contribution to the de Sitter $\mathcal{S}$-matrix by directly computing the time integral. The goal is to derive \eqref{mres}. Substituting \eqref{gsigma} to \eqref{mdef} yields
\begin{align}
&\mathcal{M}_{2\rightarrow 2} (k_{34}, k_s; \tau) = (-\tau) \int_{-\infty}^\infty \frac{\md\rho}{-2\sqrt{\pi}} e^{i \mu_\sigma \rho} \frac{1}{\sqrt{-2 k_s \tau}}  \int_{-\infty}^\infty \md(\Delta\tau)  e^{ik_{34} \Delta\tau} \nonumber\\
&\times \left[ \theta(\Delta\tau) \frac{1}{\sqrt{-i}} e^{i k_s \tau} e^{-i k_s (\tau + \Delta\tau)\cosh\rho } + \theta(-\Delta\tau) \frac{1}{\sqrt{i}} e^{-i k_s \tau} e^{i k_s (\tau + \Delta\tau) \cosh\rho} \right].
\end{align}
Performing the integral with respect to $\Delta\tau$ leads to
\begin{align}
&\mathcal{M}_{2\rightarrow 2} (k_{34}, k_s; \tau) = (-\tau) \int_{-\infty}^\infty \frac{\md\rho}{-2i\sqrt{\pi}} e^{i \mu_\sigma \rho} \frac{1}{\sqrt{-2 k_s \tau}}   \nonumber\\
&\times \left[ \frac{1}{\sqrt{i}} e^{-i k_s\tau(1- \cosh\rho)} \frac{1}{k_{34} + k_s \cosh\rho - i \varepsilon} - \frac{1}{\sqrt{-i}} e^{ik_s \tau(1-\cosh\rho)} \frac{1}{k_{34} - k_s\cosh\rho + i \varepsilon} \right]. \label{intrho}
\end{align}
Then, we have to compute the integral with respect to $\rho$, which has the form
\begin{equation}
\int_{-\infty}^\infty \md\rho ~F(\rho) e^{-ik_s\tau \alpha(\rho)} ,
\end{equation}
where $F(\rho)$ is an analytic function and
\begin{equation}
\alpha(\rho) = 1 - \cosh\rho - \frac{\mu_\sigma}{k_s\tau} \rho.
\end{equation}
Such an integral can be computed using the saddle-point approximation
\begin{equation}
\int_{-\infty}^\infty \md\rho ~F(\rho) e^{-ik_s\tau \alpha(\rho)} = F(\rho_0) e^{-ik_s\tau \alpha(\rho_0)} \sqrt{\frac{2\pi}{ -k_s\tau \abs{\alpha''(\rho_0))}}} e^{i \pi \mathrm{sgn(\alpha''(\rho_0))}/4},
\end{equation}
where $\rho = \rho_0$ is the saddle point of $\alpha(\rho)$ given by
\begin{equation}
\alpha'(\rho_0) \equiv 0 = - \sinh\rho_0 - \frac{\mu_\sigma}{k_s\tau} .
\end{equation}
At the saddle point, $\alpha(\rho)$ and $\alpha''(\rho)$ are given by
\begin{align}
& \alpha(\rho_0) = 1 - \sqrt{1 - \frac{\mu_\sigma^2}{k_s^2\tau^2}} + \frac{\mu_\sigma}{k_s\tau} \sinh^{-1} \frac{\mu_\sigma}{k_s\tau} \simeq \frac{\mu^2}{k_s^2\tau^2}, \nonumber\\
& \alpha''(\rho_0) = -\cosh\rho_0 = - \sqrt{1 - \frac{\mu_\sigma^2}{k_s^2\tau^2}} < 0.
\end{align}
Substituting them into \eqref{intrho} yields
\begin{align}
\mathcal{M}_{2\rightarrow 2} (k_{34}, k_s; \tau) &\simeq \frac{-\tau}{2 i k_s \tau} \left( \frac{1}{k_{34} + k_s \cosh\rho_0 - i \varepsilon} - \frac{1}{k_{34} - k_s \cosh\rho_0 + i\varepsilon} \right) \nonumber\\
&= \frac{-\tau}{2 i k_s \tau} \frac{- 2 k_s \cosh\rho_0}{k_{34}^2 - k_s^2 \cosh^2\rho_0 + i\varepsilon} \nonumber\\
&\simeq  \frac{i}{k_{34}^2 - k_s^2 - (\mu_\sigma/\tau)^2 + i \varepsilon}.
\end{align}

\section{Differential equation \label{diffeq}}

In this appendix, we show the technical derivation of the massive field exchange contribution to the de Sitter $\mathcal{S}$-matrix by solving a differential equation. If we define the following function
\begin{equation}
\mathscr{F}(p,q) \equiv - \int_{-\infty}^0 \frac{\md z}{z} (-z)^{\frac{d-2}{2}} \int_{-\infty}^0 \frac{\md z'}{z'} (-z')^{\frac{d-2}{2}} e^{-i z p} H G_\sigma(z,z') e^{i z' q},
\end{equation}
we can show that it satisfies the following differential equation
\begin{equation}
\left[ (p^2 - 1) \partial_p^2 + (d-1) p \partial_p + \mu_\sigma^2 + \frac{(d-2)^2}{4} \right] \mathscr{F} = \frac{i \Gamma(d-2)}{(i (q-p))^{d-2}}. \label{diffeqf}
\end{equation}
With the equation of motion of the mode function \eqref{eom}, we can show that the propagator satisfies
\begin{equation}
\left[ (z \partial_z)^2 + z^2 + \mu_\sigma^2 \right] G_\sigma(z,z') = \frac{i}{H} z \delta(z-z').
\end{equation}

To prove the differential equation, we start by defining
\begin{equation}
\mathscr{G}(p,q) \equiv - \int_{-\infty}^0 \frac{\md z}{z} (-z)^{\frac{d-2}{2}} \int_{-\infty}^0 \frac{\md z'}{z'} (-z')^{\frac{d-2}{2}} e^{-i z p} e^{i z' q} H \left[ (z \partial_z)^2 + z^2 + \mu_\sigma^2 \right] G_\sigma(z,z'),
\end{equation}
which can be easily solved to
\begin{equation}
\mathscr{G}(p,q) = \frac{i \Gamma(d-2)}{(i (q-p))^{d-2}}.
\end{equation}
With integration by parts, we obtain
\begin{align}
\mathscr{G}(p,q) =& - \int_{-\infty}^0 \md z \left\lbrace \partial_z \left[ (-z) \partial_z \left( (-z)^{\frac{d-2}{2}} e^{-ipz} \right) \right] + (z^2 + \mu_\sigma^2) e^{-ipz} \right\rbrace \nonumber\\
& \times \int_{-\infty}^0 \frac{\md z'}{-z'} (-z')^{\frac{d-2}{2}} e^{iqz'} HG_\sigma(z,z') \nonumber\\
=& - \int_{-\infty}^0 \frac{\md z}{-z}  (-z)^{\frac{d-2}{2}} \left[ \frac{d^2}{4} + 1 + i p z - d(1 + i p z) - p^2 z^2 + \mu_\sigma^2 + z^2 \right] e^{-ipz} \nonumber\\
& \times \int_{-\infty}^0 \frac{\md z'}{-z'} (-z')^{\frac{d-2}{2}} e^{iqz'} HG_\sigma(z,z') \nonumber\\
=& \left[ (p^2 - 1) \partial_p^2 + (d-1) p \partial_p + \mu_\sigma^2 + \frac{(d-2)^2}{4} \right] \mathscr{F}(p,q),
\end{align}
which proves the differential equation \eqref{diffeqf}. Then, we can connect $\mathscr{F}$ and $\mathcal{A}_{2 \rightarrow 2}'$ by identifying $p = k_{12}/k_s$ and $q = k_{34}/k_s$.

We are interested in the solution of the differential equation in the limit $E \rightarrow 0$ (or $p \rightarrow q$) and $H \rightarrow 0$, where we can substitute an ansatz $\mathscr{F} \sim (q-p)^n$. The first and second terms in the LHS of the differential equation can be approximated as $(q^2 - 1) \partial_p^2 (q-p)^n \sim (q - p)^{n-2}$ and $q \partial_p (q-p)^n \sim (q - p)^{n-1}$. Because $H \rightarrow 0$, the third term dominates the fourth term in the LHS of the differential equation  $\mu_\sigma \gg (d-2)/2$. Multiplying the third term with $\mathscr{F}$ yields $\mu^2(q-p)^n \sim (q-p)^2/H^2$, which is comparable to the first term because $q-p$ and $H$ have the same order.

\bibliographystyle{jhep}
\bibliography{Reference}

\end{document}